\documentclass[twocolumn, nofootinbib]{revtex4}
\begin{document}
\title{Spinning-particle model for the Dirac equation and the relativistic \it{\textbf{Zitterbewegung}}}

\author{A. A. Deriglazov} \email{alexei.deriglazov@ufjf.edu.br}
\altaffiliation{On leave of absence from Dep. Math. Phys., Tomsk Polytechnical University, Tomsk, Russia.}

\affiliation{Depto. de Matem\'atica, ICE, Universidade Federal de
Juiz de Fora, MG, Brazil}

\begin{abstract}
We construct the relativistic particle model without Grassmann variables which
meets the following requirements. A) Canonical quantization of the model implies the Dirac equation. B) The variable
which experiences {\it Zitterbewegung}, represents a gauge non-invariant variable in our model. Hence our particle does
not experience the undesirable {\it Zitterbewegung}. C) In the non-relativistic limit spin is described by
three-vector, as it could be expected.
\end{abstract}

\maketitle 

\section{Introduction}
The Dirac spinor $\Psi$ can be used to construct the four-dimensional current vector, $\bar\Psi\Gamma^\mu\Psi$, which
preserves for solutions to the Dirac equation, $\partial_\mu(\bar\Psi\Gamma^\mu\Psi)=0$. As a consequence, the integral
over space of its null-component, $\int d^3x\Psi^\dagger\Psi$, does non depend on time. Hence the quantity
$\Psi^\dagger\Psi\ge 0$ admits the probabilistic interpretation, and we expect that one-particle sector of the Dirac
equation can be described in the framework of relativistic quantum mechanics. Although the true understanding of spin
is achieved in the framework of quantum electrodynamics, a lot of efforts has been spent in attempts to construct
semiclassical description of relativistic spin on the base of mechanical models [1-12]. However, the relativistic
quantum mechanics applied to the one-particle solutions predicts some controversial effects, like the {\it
Zitterbewegung} of the center-of-charge position operator [1, 2] and the Klein's paradox.

Analyzing the applicability of quantum-mechanical treatment to the
free Dirac equation, Schr\"odinger noticed [1] that the
center-of-charge position operator in the Heisenberg picture
experiences rapid oscillations called {\it Zitterbewegung}. If we
take the state vector with positive and negative energy
components, the expectation value of the operator has a similar
behavior. It is often assumed that {\it Zitterbewegung} represents
the physically observable motion of a real particle [18, 19]. The
analogous systems that are described by a Dirac-type equation and
simulate {\it Zitterbewegung} are under intensive study in
different physical set-ups, including graphene, trapped ions,
photonic lattices and ultracold atoms (see [22] and the references
therein). In this work we show that the status of this phenomenon
in relativistic quantum mechanics is not as clear as is commonly
believed.

Besides the center of charge, $x$, in the Dirac theory we can construct the center-of-mass (Pryce-Newton-Wigner) [3, 4]
operator $\tilde x$ in such a way that the conjugated momentum of $x$ turns out to be the mechanical momentum for
$\tilde x$. So the Dirac particle looks like a kind of composed system (this picture has been used by Schr\"odinger [1]
to identify spin with inner angular momentum of the system). It further complicates the semiclassical analysis, as the
Dirac equation gives no evidence as regards which of these two operators should be identified with the position of the
particle.

To understand the controversial properties of the one-particle
Dirac equation, it would be desirable to have at our disposal the
semiclassical particle model which leads to the Dirac equation in
the course of canonical quantization. It implies, in particular,
that $\Gamma$\,-matrices should be produced through quantization
of some set of classical variables. While the problem has a long
history (see [3-12] and the references therein), there appears to
be no wholly satisfactory solution to date. The main difficulty
consists of the proper choice of the basic classical variables for
construction of the spin space. If we start from some
classical-mechaniccs action functional, the phase-space variables,
say $\omega^a$, $\pi_b$, necessarily obey the Poisson bracket
algebra $\{\omega^a, \pi_b\}=\delta^a{}_b$. The number of
variables and the algebra are different from the number of spin
operators and their commutators (for instance, for
non-relativistic spin they are $[\hat S_i, \hat
S_j]=i\hbar\epsilon_{ijk}\hat S_k)$. To improve this, we need to
impose constraints as well as to pass from the initial to some
composed variables. This implies the use of the Dirac machinery
for constrained theories [13-16]. Although the non-relativistic
spin can be described along these lines [17], it seems to be
surprisingly difficult\footnote{The Berezin-Marinov model [9] is
based on the Grassmann variables and leads to the Dirac equation.
The problem here is that the Grassmann classical mechanics
represents a rather formal mathematical construction. It leads to
certain difficulties [9, 14] in attempts to use it for description
of the spin effects on the semiclassical level, before the
quantization. The Hanson-Regge theory [8] does not produce
$\Gamma$-matrices. The Barut-Zanghi model [10] does not imply the
Dirac equation.} to construct, in a systematic way, a consistent
model for the relativistic spin.

In this letter we propose the relativistic particle model without Grassmann (anticommutative) variables which leads to
the Dirac equation. We apply our model to analysis of the relativistic {\it Zitterbewegung}, presenting a simple
semiclassical argument on the non-physical character of this phenomenon, which in our model represents the dynamics of
an unobservable variable.

\par
\noindent
\section{Model-independent construction of the relativistic spin surface}
To construct the relativistic spin surface, we start from the Dirac equation
\begin{eqnarray}\label{1.1}
\left[\Gamma^\mu(\hat p_\mu+\frac{e}{c}A_\mu)+mc\right]\Psi=0.
\end{eqnarray}
Applying the operator $\Gamma^\mu(\hat p_\mu+\frac{e}{c}A_\mu)-mc$, this implies the Klein-Gordon equation with
non-minimal interaction
\begin{eqnarray}\label{1.1_1}
\left[(\hat p^\mu+\frac{e}{c}A^\mu)^2+\frac{e\hbar}{2c}F_{\mu\nu}\Gamma^{\mu\nu}+m^2c^2\right]\Psi=0,
\end{eqnarray}
where
\begin{eqnarray}\label{1.2}
\Gamma^{\mu\nu}\equiv\frac{i}{2}(\Gamma^\mu\Gamma^\nu-\Gamma^\nu\Gamma^\mu),
\end{eqnarray}
and $F_{\mu\nu}=\partial_\mu A_\nu-\partial_\nu A_\mu$. We use the representation with hermitean $\Gamma^0$ and
antihermitean $\Gamma^i$
\begin{eqnarray}\label{1.1_2}
\Gamma^0=
\left(
\begin{array}{cc}
1& 0\\
0& -1
\end{array}
\right), \quad
\Gamma^i=
\left(
\begin{array}{cc}
0& \sigma^i\\
-\sigma^i& 0
\end{array}
\right),
\end{eqnarray}
then $[\Gamma^\mu, \Gamma^\nu]_{+}=-2\eta^{\mu\nu}$,
$\eta^{\mu\nu}=(- + + +)$, and $\Gamma^0\Gamma^i$, $\Gamma^0$ are
the Dirac matrices $\alpha^i$, $\beta$ [2]. We take the classical
counterparts of the operators $\hat x^\mu$ and $\hat
p_\mu=-i\hbar\partial_\mu$ in the standard way, which are $x^\mu$,
$p^\nu$, with the Poisson brackets $\{x^\mu,
p^\nu\}=\eta^{\mu\nu}$.

Let us discuss the classical variables that could produce the
$\Gamma$\,-matrices. The relativistic equation for the spin
precession is usually obtained including the three-dimensional
spin vector $S^i$ either into the Frenkel tensor $\Phi^{\mu\nu}$,
$\Phi^{\mu\nu}u_\nu=0$, or into the Bargmann-Michel-Telegdi
four-vector\footnote{The conditions $\Phi^{\mu\nu}u_\nu=0$ and
$S^\mu u_\mu=0$ guarantee that in the rest frame only three
components of these quantities survive, which implies the right
non-relativistic limit.} $S^\mu$, $S^\mu u_\mu=0$, where $u_\nu$
represents four-velocity of the particle. However, the
semiclassical models based on these schemes do not lead to a
reasonable quantum theory, as they do not produce the Dirac
equation through canonical quantization. We now present arguments
as to how this can be achieved in the formulation that implies
inclusion of $S^i$ into the $SO(2, 3)$ angular momentum tensor
$J^{AB}$ of five-dimensional space, $A=(\mu, 5)=(0, 1, 2, 3, 5)$,
with the metric $\eta^{AB}=(- + + + -)$.

First, we analyze commutators of the $\Gamma$\,-matrices. The commutators do not form closed Lie algebra, but produce
$SO(1, 3)$\,-Lorentz generators (\ref{1.2}). The set $\Gamma^\mu$,  $\Gamma^{\mu\nu}$ forms closed algebra
\begin{eqnarray}\label{1.3}
[\Gamma^\mu, \Gamma^\nu]=-2i\Gamma^{\mu\nu}, \quad
[\Gamma^{\mu\nu},
\Gamma^\alpha]=2i(\eta^{\mu\alpha}\Gamma^\nu-\eta^{\nu\alpha}\Gamma^\mu), ~ \cr
[\Gamma^{\mu\nu}, \Gamma^{\alpha\beta}]=
2i(\eta^{\mu\alpha}\Gamma^{\nu\beta}-
\eta^{\mu\beta}\Gamma^{\nu\alpha}-
\eta^{\nu\alpha}\Gamma^{\mu\beta}+\eta^{\nu\beta}\Gamma^{\mu\alpha}).
\end{eqnarray}
The algebra can be identified with the five-dimensional Lorentz
algebra  $SO(2, 3)$ with generators $\hat J^{AB}$
\begin{eqnarray}\label{1.4}
[\hat J^{AB}, \hat J^{CD}]=2i(\eta^{AC}\hat J^{BD}-\eta^{AD}\hat J^{BC}- \cr
\eta^{BC}\hat J^{AD}+\eta^{BD}\hat J^{AC}),
\end{eqnarray}
assuming $\Gamma^\mu\equiv \hat J^{5\mu}$, $\Gamma^{\mu\nu}\equiv
\hat J^{\mu\nu}$.

To reach the algebra starting from a classical-mechanics model, we
introduce ten-dimensional "phase" space of the spin degrees of
freedom, $\omega^A$, $\pi^B$, equipped with the Poisson bracket
$\{\omega^A, \pi^B\}=\eta^{AB}$.
Consider the inner angular momentum
\begin{eqnarray}\label{1.6}
J^{AB}\equiv 2(\omega^A\pi^B-\omega^B\pi^A).
\end{eqnarray}
Poisson brackets of these quantities form the algebra
\begin{eqnarray}\label{1.4.1}
\{J^{AB}, J^{CD}\}_{PB}=2(\eta^{AC} J^{BD}-\eta^{AD}J^{BC}- \cr \eta^{BC}J^{AD}+\eta^{BD}J^{AC}).
\end{eqnarray}
Comparing (\ref{1.4.1}) with (\ref{1.4}) we conclude that the operators $\Gamma^\mu$, $\Gamma^{\mu\nu}$ could be
obtained by quantization of $J^{AB}$.

Since $J^{AB}$ are the variables which we are interested in, we try to take them as coordinates of the space $\omega^A,
\pi^B$. The Jacobian of the transformation $(\omega^A, \pi^B)\rightarrow J^{AB}$ has rank equal seven\footnote{The rank
has been computed using the program: Wolfram Mathematica 8.}. So, only seven among ten functions $J^{AB}(\omega, \pi)$,
$A<B$, are independent quantities. They can be separated as follows. By construction, the quantities (\ref{1.6}) obey
the identity $\epsilon^{\mu\nu\alpha\beta}J^5{}_\nu J_{\alpha\beta}=0$, this can be solved as
\begin{eqnarray}\label{1.61}
J^{ij}=(J^{50})^{-1}(J^{5i}J^{0j}-J^{5j}J^{0i}).
\end{eqnarray}
Hence we can take $J^{5\mu}$, $J^{0i}$ as the independent variables. We could complete the set up to a base of the
phase space $(\omega^A, \pi^B)$ adding three more coordinates, for instance $\omega^3$, $\omega^5$, $\pi^5$. Quantizing
the complete set we obtain, besides the desired operators $\hat J^{5\mu}, \hat J^{0i}$, some extra operators
$\hat\omega^3$, $\hat\omega^5$, $\hat\pi^5$. They are not present in the Dirac theory, and are not necessary for
description of spin. So we need to reduce the dimension of our space from ten to seven imposing three constraints.
There is one important restriction on the choice of constraints. Canonical quantization of a system with constraints
implies replacement the Poisson by the Dirac bracket, the latter is constructed with help of the constraints. We need
$SO(2, 3)$\,-invariant constraints $T_a$, $\{T_a, J^{AB}\}_{PB}=0$, otherwise the Dirac-brackets algebra will not
coincide with those of the Poisson, (\ref{1.4.1}).

The only quadratic $SO(2, 3)$\,-invariants which can be constructed from $\omega^A$, $\pi^B$ are $\omega^A\omega_A$,
$\omega^A\pi_A$ and $\pi^A\pi_A$. So we restrict our model to live on the surface defined by the equations
\begin{eqnarray}\label{1.7}
T_3\equiv\pi^A\pi_A+a_3=0;
\end{eqnarray}
\begin{eqnarray}\label{1.71}
T_4\equiv\omega^A\omega_A+a_4=0, \quad T_5\equiv\omega^A\pi_A=0,
\end{eqnarray}
where $a_3$, $a_4$ are some numbers.

The matrix $\frac{\partial(J^{5\mu}, J^{0i}, T_4, T_5, \omega^5)}{\partial(\omega^A, \pi^B)}$ has rank equal ten. So
the quantities
\begin{eqnarray}\label{1.71.1}
J^{5\mu}, ~ J^{0i}, ~ T_4, ~ T_5, ~ \omega^5,
\end{eqnarray}
can be taken as coordinates of the space $(\omega^A$, $\pi^B)$.
The equation $J^{AB}=2(\omega^A\pi^B-\omega^B\pi^A)$ implies the
identity
\begin{eqnarray}\label{1.68}
J^{AB}J_{AB}=8[(\omega^A)^2(\pi^B)^2-(\omega^A\pi_A)^2]= \cr
8[(T_4-a_4)(T_3-a_3)-(T_5)^2], \quad
\end{eqnarray}
then the constraint $T_3$ can be written in the coordinates
(\ref{1.71.1}) as follows:
\begin{eqnarray}\label{1.71.2}
T_3=\frac{(J^{AB})^2+8(T_5)^2}{8(T_4-a_4)}+a_3,
\end{eqnarray}
where $J^{ij}$ are given by Eq. (\ref{1.61}). Note that $T_3$ does not depend on $\omega^5$. On the hyperplane
$T_4=T_5=0$ it reduces to
\begin{eqnarray}\label{1.71.3}
-8a_4T_3=(J^{AB})^2-8a_3a_4=0.
\end{eqnarray}
Eq. (\ref{1.71.3}) states that the value of $SO(2, 3)$\,-Casimir
operator\footnote{In quantum theory, for the operators
(\ref{1.4}), (\ref{1.3}) we have: $\hat J^{AB}\hat
J_{AB}=20\hbar^2$.} $(J^{AB})^2$ is equal to $8a_3a_4$.

In the dynamical model constructed below, the equation $T_3=0$ appears as the first-class constraint. It implies that
we deal with a theory with local symmetry, with the constraint being the generator of the symmetry [20]. The coordinate
$\omega^5$ is not inert under the symmetry, $\delta\omega^5\sim\{T_3, \omega^5\}\ne 0$. Hence $\omega^5$ is gauge
non-invariant variable.

Summing up, we have restricted dynamics of spin on the surface (\ref{1.7}), (\ref{1.71}). If (\ref{1.71.1}) are taken
as coordinates of the phase space, the surface is the hyperplane $T_4=T_5=0$ with the coordinates $J^{5\mu}, J^{0i},
\omega^5$ subject to the condition (\ref{1.71.3}). Since $\omega^5$ is gauge non-invariant coordinate, we can discard
it. It implies that we can quantize $J^{5\mu}, J^{0i}$ instead of the initial variables $\omega^A$, $\pi^B$.

Following the canonical quantization paradigm, the variables must
be replaced by Hermitian operators\footnote{The matrices
$\Gamma^\mu$, $\Gamma^{\mu\nu}$ are Hermitian operators with
respect to the scalar product $(\Psi_1,
\Psi_2)=\Psi_1^\dagger\Gamma^0\Psi_2$.} with commutators
resembling the Poisson bracket
\begin{eqnarray}\label{1.71.5}
[ ~ , ~ ]=i\hbar\left.\{ ~  ,  ~ \}\right|_{J\rightarrow\hat J}.
\end{eqnarray}
Similarly to the case of $\Gamma$\,-matrices, brackets of the
variables $J^{5\mu}$, $J^{0i}$ do not form closed Lie algebra. The
non closed brackets are
\begin{eqnarray}\label{1.71.6}
\{J^{5i}, J^{5j}\}=\{J^{0i}, J^{0j}\}=-2J^{ij},
\end{eqnarray}
where $J^{ij}$ is given by Eq. (\ref{1.61}). Adding them to the
initial variables, we obtain the set $J^{AB}=(J^{5\mu}, J^{0i},
J^{ij})$ which obeys the desired algebra (\ref{1.4.1}).

According to Eqs. (\ref{1.4}), (\ref{1.4.1}) the quantization is achieved replacing the classical variables $J^{5\mu}$,
$J^{\mu\nu}$ on $\Gamma$\,-matrices\footnote{Replacing (\ref{1.61}) by an operator $\hat J^{ij}(\Gamma^\mu,
\Gamma^{0i})$ we arrange the operators $\Gamma$ in such a way, that $\hat J^{ij}(\Gamma)=\Gamma^{ij}$.}.  We assume
that $\omega^A$ has a dimension of length, then $J^{AB}$ has the dimension of the Planck's constant. Hence the
quantization rule is
\begin{eqnarray}\label{1.9}
J^{5\mu}\rightarrow\hbar\Gamma^\mu, \quad
J^{\mu\nu}\rightarrow\hbar\Gamma^{\mu\nu}.
\end{eqnarray}

This implies that the Dirac equation (\ref{1.1}) can be produced
by the constraint (since the {\it Zitterbewegung} is a property of
the free Dirac equation, we take temporarily $A_\mu=0$)
\begin{eqnarray}\label{1.8}
T_2\equiv p_\mu J^{5\mu}+mc\hbar=0.
\end{eqnarray}
In quantum theory, the Dirac equation implies the Klein-Gordon
one. In contrast, in the classical theory the constraint
(\ref{1.8}) does not imply the mass-shell condition
\begin{eqnarray}\label{1.10}
T_1\equiv p^2+m^2c^2=0.
\end{eqnarray}
To improve this, we are forced to look for a classical model that produces this equation as an independent constraint.
The model without the constraint (\ref{1.10}) has been considered in [21]. It shows the same undesirable properties as
those of Dirac equation in the classical limit.

\par
\noindent
\section{Spinning-particle action, canonical quantization and the Dirac equation}
According to the previous section, to describe the relativistic spin we need a theory that implies the Dirac constraints (\ref{1.7}), (\ref{1.71}), (\ref{1.8}), (\ref{1.10}).

We recall that the Hamiltonian action for a system with the
phase-space variables $Q^\alpha$, $P_\alpha$ reads $S=P_\alpha\dot
Q^\alpha-H$, $H=H_0+\lambda_a\Phi_a$, where $H_0$ is the canonical
Hamiltonian and $\lambda_a$ are the Lagrangian multipliers for the
primary constraints $\Phi_a$. For the present case, we propose to
consider the total Hamiltonian of the form
$H=\frac12e_aT_a+\lambda_{ea}\pi_{ea}$, where $\pi_{ea}$ are
conjugate momenta for the auxiliary variables $e_a$, the latter
are associated with the constraints $T_a$, $a=1, 2, 3, 4$. So, let
us consider the following $d=4$ Poincar\'e invariant action
\begin{eqnarray}\label{Z2.1}
S=\int d\tau \, p_\mu\dot x^\mu+\pi_A\dot\omega^A-\left[\frac{e_1}{2}(p^2+m^2c^2)+\right. \qquad \cr
\left.\frac{e_2}{2}(p_\mu J^{5\mu}+mc\hbar)+\frac{e_3}{2}T_3+\frac{e_4}{2}T_4+\pi_{ea}(\lambda_{ea}-\dot e_a)\right],
\end{eqnarray}
If we omit the spin-space coordinates, $\omega^A=\pi_A=0$, Eq. (\ref{Z2.1}) reduces to the well known action of a spinless relativistic particle
\begin{eqnarray}\label{Z2.1_0}
S_0=\int d\tau p_\mu\dot x^\mu-\frac12e(p^2+m^2c^2)+\pi_e(\dot e-\lambda_e).
\end{eqnarray}

Variation of the action (\ref{Z2.1}) with respect to $e_a$ leads to the constraints $T_a=0$. Preservation in time of
the constraint $T_4$, $\dot T_4=0$, implies $T_5=0$, that is we reproduce all the desired constraints (\ref{1.7}),
(\ref{1.71}), (\ref{1.8}), (\ref{1.10}). The constraints $T_1$, $T_2$ have vanishing Poisson brackets with all the
constraints. The remaining constraints obey the Poisson-bracket algebra
\begin{eqnarray}\label{Z2.1_02}
\{T_3, T_4\}=-4T_5, \qquad \{T_3, T_5\}=-2T_3+2a_3, \cr \{T_4,
T_5\}=2T_4-2a_4. \qquad \qquad \qquad
\end{eqnarray}
If we take the combination
\begin{eqnarray}\label{Z2.1_03}
\tilde T_3\equiv T_3+\frac{a_3}{a_4}T_4,
\end{eqnarray}
the algebra acquires the form
\begin{eqnarray}\label{Z2.1_01}
\{\tilde T_3, T_4\}=-4T_5, \quad \{\tilde T_3,
T_5\}=-2T_3+2\frac{a_3}{a_4}T_4, \cr \{T_4, T_5\}=2T_4-2a_4.
\qquad \qquad
\end{eqnarray}
The only bracket which does not vanish on the constraint surface
is $\{T_4, T_5\}$.  According the Dirac terminology [13-15], we
have the first-class constraints (\ref{Z2.1_03}), (\ref{1.8}),
(\ref{1.10}), and the second-class pair (\ref{1.71}). The presence
of the first-class constraints indicates that we are dealing with
a theory invariant under a three-parameter group of local (gauge)
symmetries, which will be found below.

The constraints (\ref{Z2.1_03}), (\ref{1.71}) can be taken into
account by transition from the Poisson to the Dirac bracket, and
after that they can be omitted from consideration [13-15]. Since
the Dirac brackets are constructed with use of $SO(2,
3)$\,-invariants, the Dirac brackets of the quantities $J^{AB}$
coincide with the Poisson one (\ref{1.4.1}). Hence we quantize the
model according Eq. (\ref{1.9}). The operator produced by the
first-class constraint (\ref{1.8}) is imposed on the state vector.
This gives the Dirac equation. In the result, canonical
quantization of the model leads to the desired quantum picture.

The Lagrangian formulation can be restored from Eq. (\ref{Z2.1}).
To achieve this, we note that the conjugate momenta enter into the
Hamiltonian in the form $\frac12 P_\alpha G^{\alpha\beta}P_\beta$,
where $P_\alpha\equiv(p_\mu, \pi_\nu, \pi^5)$ and the "metric" ~
$G^{\alpha\beta}(\omega^\mu, \omega^5, e_1, e_2, e_3)$ is given by
$9\times 9$\,-matrix
\begin{eqnarray}
\left(
\begin{array}{ccccccccc}
{}& {}& {}& |& {}& {}& {}& |& -e_2\omega^0\\
{}& e_1\eta& {}& |& {}& e_2\omega^5\eta& {}& |& \ldots\\
{}& {}& {}& |& {}& {}& {}& |& -e_2\omega^3\\
\mbox{---}& \mbox{---}& \mbox{---}& {}& \mbox{---}& \mbox{---}& \mbox{---}& {}& \mbox{---}\\
{}& {}& {}& |& {}& {}& {}& |& 0\\
{}& e_2\omega^5\eta& {}& |& {}& e_3\eta& {}& |& \ldots\\
{}& {}& {}& |& {}& {}& {}& |& 0\\
\mbox{---}& \mbox{---}& \mbox{---}& {}& \mbox{---}& \mbox{---}& \mbox{---}& {}& \mbox{---}\\
-e_2\omega^0& \ldots& -e_2\omega^3& |& 0& \ldots& 0& |& -e_3
\end{array}
\right) \nonumber
\end{eqnarray}
Here $\eta$ is the Minkowski metric. To find the Lagrangian, we
solve the Hamiltonian equations for the position variables
$Q^\alpha\equiv(x^\mu, \omega^\nu, \omega^5)$, $\dot
Q^\alpha=G^{\alpha\beta}P_\beta$, with respect to $P_\alpha$. It
gives $P_\alpha=G_{\alpha\beta}\dot Q^\beta$, where
$G_{\alpha\beta}$ is the inverse metric. We substitute these
$P_\alpha$ back into Eq. (\ref{Z2.1}), obtaining the Lagrangian
\begin{eqnarray}\label{Z2.1_2}
L=\frac12G_{\alpha\beta}\dot Q^\alpha\dot
Q^\beta-\frac12e_4\omega^A\omega_A-\frac12e_ba_b.
\end{eqnarray}
It has been denoted $(a_1, a_2, a_3, a_4)=(m^2c^2, mc\hbar, a_3,
a_4)$. The kinetic term looks like those of a free particle moving
on the curved nine-dimensional space with the metric
$G_{\alpha\beta}$.

We introduce the abbreviation
\begin{eqnarray}\label{Z2.9}
Dx^\mu\equiv\dot x^\mu-\frac{e_2}{e_3}J_{L}^{5\mu},
\end{eqnarray}
where
$J_{L}^{5\mu}\equiv\omega^5\dot\omega^\mu-\omega^\mu\dot\omega^5$
is the configuration-space counterpart of the angular momentum
$J^{5\mu}$. Then manifest form of the Lagrangian (\ref{Z2.1_2}) is
\begin{eqnarray}\label{Z2.10}
2L=\frac{e_3}{B}\left[(Dx^\mu)^2-\frac{e_2^2}{A}(Dx^\mu\omega_\mu)^2\right]+
\cr
\frac{1}{e_3}\dot\omega^A\dot\omega_A-e_4\omega^A\omega_A-e_ba_b.
\qquad
\end{eqnarray}
It has been denoted $B=e_1e_3-e_2^2(\omega^5)^2$,
$A=B+e_2^2(\omega^\mu)^2$. We have verified that hamiltonization
of the Lagrangian leads back to the Hamiltonian action
(\ref{Z2.1}).

\par
\noindent
\section{Classical equations of motion and the lack of \it{\textbf{Zitterbewegung}}}
In this section we show that the status of {\it Zitterbewegung}
phenomenon in relativistic quantum mechanics is not as clear as is
commonly believed.

Besides the constraints discussed above, the action (\ref{Z2.1}) implies the following equations
(we use the notation $(p\omega)=p^\mu\omega_\mu$\,)
\begin{eqnarray}\label{Z2.2}
\dot e_a=\lambda_{ea}, ~ \pi_{ea}=0, ~ a=1, 2, 3; \quad \cr
e_4=\frac{a_3}{a_4}e_3, ~ \pi_{e_4}=0, ~
\lambda_{e4}=\frac{a_3}{a_4}\lambda_{e3};
\end{eqnarray}
\begin{eqnarray}\label{Z2.3}
\dot x^\mu=e_1p^\mu+\frac12e_2J^{5\mu}, \quad \dot p^\mu=0; \quad \qquad \quad \cr
\dot\omega^\mu=e_3\pi^\mu+e_2\omega^5p^\mu, \qquad
\dot\pi^\mu=e_2\pi^5p^\mu-\frac{a_3}{a_4}e_3\omega^\mu; \cr
\dot\omega^5=e_3\pi^5+e_2(p\omega), \qquad ~
\dot\pi^5=e_2(p\pi)-\frac{a_3}{a_4}e_3\omega^5. ~
\end{eqnarray}
As a consequence, we obtain (assuming the gauge conditions
$e_a=\mbox{const}$) $\ddot x^\mu=-\frac12e_2^2J^{\mu\nu}p_\nu$,
that is $x^\mu\;$-variable is under {\it Zitterbewegung} in
non-interacting theory.

Since we deal with a constrained theory, our first task is to
specify the physical-sector variables. We note that Eqs.
(\ref{Z2.2}), (\ref{Z2.3}) do not determine the Lagrangian
multipliers $\lambda_a(\tau)$, $a=1, 2, 3$. Eqs. (\ref{Z2.2}) then
imply that $e_a$, $a=1, 2, 3$, remain an arbitrary functions as
well. The arbitrary functions enter into solutions to equations of
motion for the variables $x^\mu, \omega^A, \pi^A$. Hence, except
$p^\mu$, all the variables has ambiguous dynamics. According to
the general theory [13-15], variables with ambiguous dynamics do
not represent the observable quantities. In particular, the
coordinate $x^\mu$, which corresponds to the center-of-charge
position operator of the Dirac equation, and experiences {\it
Zitterbewegung}, {\it is an unobservable quantity in our model}.

Let us compute the total number of physical degrees of freedom.
Omitting the auxiliary variables and the corresponding
constraints, we have $18$ phase-space variables $x^\mu$,
$p_\mu$, $\omega^A$, $\pi_A$ subject to the constraints (\ref{1.7}), (\ref{1.71}), (\ref{1.8}), (\ref{1.10}).
Taking into account that each
second-class constraint rules out one variable, whereas each
first-class constraint rules out two variables, the number of
physical degrees of freedom is $18-(2+2\times 3)=10$.
To construct the unambiguous position variable, we note that the quantity
\begin{eqnarray}\label{Z2.5}
\tilde x^\mu=x^\mu+\frac{1}{2p^2}J^{\mu\nu}p_\nu,
\end{eqnarray}
obeys $\dot{\tilde x}^\mu=\tilde ep^\mu$, where $\tilde e\equiv
e_1+\frac{\hbar e_2}{2mc}$. Besides, we know $\dot p_\mu=0$. Since
these equations resemble those for a spinless relativistic
particle, the remaining ambiguity due to $\tilde e$ has the
well-known interpretation, being related with reparametrization
invariance of the theory. In accordance with this, we assume that
$\tilde x^\mu(\tau)$ represent the parametric equations of the
trajectory $\tilde x^i(t)$. Using the identity
$\frac{dA(t)}{dt}=c\frac{\dot A(\tau)}{\dot x^0(\tau)}$, we
conclude that the reparametrization-invariant variable $\tilde
x^i(t)$ has deterministic evolution: $\frac{d\tilde
x^i}{dt}=c\frac{\dot{\tilde x}^i} {\dot{\tilde
x}^0}=\frac{cp^i}{p^0}$. In the absence of interaction, it moves
along a straight line. We also notice that $\tilde x^\mu$
represents the center-of-mass (Pryce-Newton-Wigner) coordinate [3,
4], and $p_\mu$ represents its mechanical momentum. Hence the
mass-shell condition (\ref{1.10}) guarantees that the $\tilde
x^i$-particle cannot exceed the speed of light.

As the classical four-dimensional spin vector, we take the Pauli-Lubanski vector
$S^\mu=\frac12\epsilon^{\mu\nu\alpha\beta}p_\nu J_{\alpha\beta}$. It has no precession in the free theory, $\dot
S^\mu=0$. In the rest frame $p^\mu=(mc, 0, 0, 0)$, it reduces to the three-dimensional rotation generator, $S^0=0$,
$S^i=\frac12mc\epsilon^{ijk}S_{jk}$, as is expected in the non-relativistic limit. Hence the ten variables $x^i(t)$,
$p^i(t)$, $S^\mu(t)$ can be taken as the physical variables.

We have specified the physical sector from analysis of equations of motion. The more traditional way to do this
consists of analysis of local symmetries of the formulation. The action (\ref{Z2.1}) is invariant under a
three-parameter group of local symmetries. One of them is the reparametrization symmetry which we take in the form
\begin{eqnarray}\label{Z2.6}
\delta_\alpha Z=\alpha\{Z, H\}, \qquad \delta_\alpha e_a=(\alpha
e_a)\dot{}.
\end{eqnarray}
Here $Z=(x^\mu, p^\mu, \omega^A, \pi^A)$. This form of
reparametrization symmetry can be obtained as follows. It is
sufficient to consider the spinless particle action
(\ref{Z2.1_0}). The standard form of reparametrization symmetry
reads $\delta x^\mu=\alpha\dot x^\mu$, $\delta p^\mu=\alpha\dot
p^\mu$, $\delta e=(\alpha e)\dot{}$ (we have omitted the
transformations $\delta\lambda_e=(\delta e)\dot{}$ and
$\delta\pi_e=0$ which are not relevant for our discussion). We
rewrite the symmetry in equivalent form, without derivatives
acting on $x$ and $p$. Every Hamiltonian action has trivial
symmetries [20], in this case they are $\delta x^\mu=\epsilon(\dot
x^\mu-\{x^\mu, H\})$, $\delta p^\mu=\epsilon(\dot p^\mu-\{p^\mu,
H\})$, where $H$ stands for the complete Hamiltonian. Taking the
combination $\delta_\alpha+\delta_{\epsilon}$ with
$\epsilon=-\alpha$, the reparametrization invariance acquires the
form $\delta x^\mu=\alpha\{x^\mu, H\}$, $\delta
p^\mu=\alpha\{p^\mu, H\})$, $\delta e=(\alpha e)\dot{}$.

For our case (\ref{Z2.1}), the manifest form of the reparametrization symmetry is
\begin{eqnarray}\label{Z2.6.1}
\delta x^\mu=\alpha(e_1p^\mu+\frac12e_2J^{5\mu}), \qquad \delta
p^\mu=0,\qquad \delta e_a=(\alpha e_a)\dot{}, \cr
\delta\omega^\mu=\alpha(e_3\pi^\mu+e_2\omega^5p^\mu), \quad
\delta\pi^\mu=\alpha(e_2\pi^5p^\mu-\frac{a_3}{a_4}e_3\omega^\mu),
\cr \delta\omega^5=\alpha(e_3\pi^5+e_2(p\omega)), \quad ~
\delta\pi^5=\alpha(e_2(p\pi)-\frac{a_3}{a_4}e_3\omega^5).
\nonumber
\end{eqnarray}
Besides, there are two more symmetries with the local parameters $\beta(\tau)$, $\gamma(\tau)$
\begin{eqnarray}\label{Z2.7}
\delta_\beta x^\mu=\beta p^\mu, \quad \delta_\beta e_1=\dot\beta;
\end{eqnarray}
\begin{eqnarray}\label{Z2.8}
\delta_\gamma\omega^A=\gamma\pi^A, \quad \delta_\gamma e_3=\dot\gamma.
\end{eqnarray}
All the initial variables, except $p_\mu$, are not gauge invariant. The center of mass $\tilde x^\mu$ turns out to be
invariant with respect to two symmetries: $\delta_\gamma\tilde x^\mu=0$, $\delta_{\tilde\beta}\tilde x^\mu=0$, where
$\delta_{\tilde\beta}=\delta_\beta+\delta_\alpha$, $\alpha=-\frac{2p^2\beta}{2p^2e_1+e_2(pJ^5)}$. As expected,
$\tilde x^\mu$ is affected only by reparametrizations: $\delta_\alpha\tilde x^\mu=\alpha(e_1+\frac{e_2(pJ^5)}{2p^2})p^\mu$.
The spin vector $S^\mu$ is invariant under all the transformations. So we have confirmed our previous result:
$\tilde x^\mu$, $p_\mu$ and $S^\mu$ can be taken as the physical-sector variables.

We finish with a preliminary comment on interaction with an external electromagnetic field. The classical constraints
that produce Eqs. (\ref{1.1}), (\ref{1.1_1}) are $\tilde T_2\equiv(p_\mu+\frac{e}{c}A_\mu)J^{5\mu}+mc\hbar=0$, $\tilde
T_1\equiv(p^\mu+\frac{e}{c}A^\mu)^2+\frac{e}{2c}F_{\mu\nu}J^{\mu\nu}+m^2c^2=0$. Their Poisson bracket reads $\{\tilde
T_2, \tilde T_1\}=-\frac{e}{2c}\partial_\alpha F_{\mu\nu}J^{5\alpha}J^{\mu\nu}$. For the homogeneous electric and
magnetic fields, the constraints form the first-class system, $\{\tilde T_2, \tilde T_1\}=0$. Hence the interaction
does not break the local symmetries presented in our model. In the general case, the breaking is in a sense "soft",
i.e., proportional to $\hbar^2$. Hence, to construct an interaction with an arbitrary field, one can start the
iteration procedure, adding the non-minimal-interaction terms of order $\hbar^2$ or more to the constraints $\tilde
T_1$, $\tilde T_2$.

\par
\noindent
\section{Conclusion}
In this work we have constructed a semiclassical model
(\ref{Z2.1}), (\ref{Z2.10}) for description of the relativistic
spin and showed its consistency both on the classical and on the
quantum level. Canonical quantization of the model leads to the
Dirac equation. As we could expect for the relativistic spinning
particle, the physical sector of the model is composed by the
position variable $\tilde x^\mu$ (\ref{Z2.5}) and the spin vector
$S^\mu$. In the absence of interaction, they obey the free
equations $\ddot{\tilde x}^\mu=0$, $\dot S^\mu=0$.

We have presented a simple semiclassical argument that prohibits
the relativistic {\it Zitterbewegung}. Roughly speaking, the
argument is as follows. The Dirac equation (\ref{1.1}) implies the
Klein-Gordon one (\ref{1.1_1}). In contrast, in the classical
theory the corresponding constraint (\ref{1.8}) does not imply the
mass-shell constraint $p^2+m^2c^2=0$. To obtain a consistent
picture, we are forced to construct the semiclassical model that
produces both constraints. In turn, the presence of independent
constraints implies that we deal with a theory with the local
symmetries (\ref{Z2.6})-(\ref{Z2.8}). Physical quantities are
those invariant under the local symmetries. Our observation is
that the classical variable $x^\mu$, that corresponds to the
center-of-charge position operator of the Dirac theory, and
experiences {\it Zitterbewegung}, is not invariant. Similarly to
the potential $A^\mu$ of electromagnetic field, $x^\mu$ is an
unobservable quantity.

\section{Acknowledgments}
This work has been supported by the Brazilian foundation FAPEMIG.

\end{document}